\documentclass[reprint,amsmath,amssymb,aps,superscriptaddress,nofootinbib,twocolumn,10pt]{revtex4-1}
\usepackage[colorlinks,linkcolor=blue,citecolor=blue,urlcolor=blue]{hyperref}
\usepackage{graphicx}
\usepackage{dcolumn}
\usepackage{subfigure}
\usepackage{bm}
\usepackage{amsmath,amsfonts,amssymb,graphicx,multirow,dcolumn,bm,latexsym,soul,nicefrac}
\usepackage{acronym}
\usepackage{enumitem}
\usepackage{array}
\usepackage{multirow}
\usepackage{appendix}
\usepackage{longtable}
\usepackage{footnote}
\usepackage{amssymb}
\usepackage{xcolor}

\newcommand{\TRC}{MOE Key Laboratory of TianQin Mission, TianQin Research Center for
Gravitational Physics \& School of Physics and Astronomy, Frontiers
Science Center for TianQin, Gravitational Wave Research Center of
CNSA, Sun Yat-sen University (Zhuhai Campus), Zhuhai 519082, China.}

\begin{document}

\title{Probing the Spin-Induced Quadrupole Moment of Massive Black Holes\\ with the Inspiral of Binary    Black Holes}

\author{Ying-Lin Kong}
\affiliation{\TRC}
\author{Jian-dong Zhang}
 \email{zhangjd9@mail.sysu.edu.cn}
\affiliation{\TRC}

\newacro{GR}{general relativity}
\newacro{GW}{gravitational wave}
\newacro{MBHB}{massive black hole binary}
\newacro{EMRI}{extreme mass-ratio inspiral}
\newacro{BH}{black hole}
\newacro{NS}{neutron star}
\newacro{BS}{boson star}
\newacro{ECO}{exotic compact object}
\newacro{LVK}{LIGO-Virgo-KAGRA}
\newacro{BBH}{binary black hole}
\newacro{BNS}{binary neutron star}
\newacro{NSBH}{neutron star-black hole}
\newacro{SIQM}{spin induced quadrupole moment}
\newacro{ASD}{amplitude spectral density}
\newacro{PSD}{power spectral density}
\newacro{SNR}{signal-to-noise ratio}
\newacro{PE}{parameter estimation}
\newacro{PN}{post-Newtonian}
\newacro{FIM}{Fisher information matrix}
\newacro{TDI}{time delay interferometry}
\newacro{ISCO}{innermost stable circular orbit}
\newacro{PDF}{probability distribution function}

\def\ccr#1{{\color{red} #1}}

\begin{abstract}
One of the most important sources for space-borne gravitational wave detectors such as TianQin and LISA, is the merger of massivie black hole binaries.
By analyzing the inspiral signals, we can probe the characteristics of massive black holes, including the spin-induced multipole moments.
By verifying the relation among mass, spin, and quadrupole moment, the no-hair theorem can be tested.
In this work, we analyed the capability of probing the spin-induced quadrupole moment with the inspiral signal of massive black hole binaries using space-borne gravitational wave detectors.
Using the Fisher information matrix, we find that the deviation of the quadrupole moment can be constrained to the level of $10^{-1}$, and events with higher mass ratios will provide a better constraint.
We also find that the late inspiral part will dominate the result of parameter estimation.
The results of Bayesian analysis indicate that the capability will be significantly enhanced by considering higher modes.
We also calculate the Bayes factor, and the results indicate that the model of a black hole and a Boson star can be distinguished without a doubt.
\end{abstract}

\maketitle

\section{Introduction}\label{sec:intro}
After the first detection of the \ac{GW} from GW150914 \cite{abbott2016gw150914},
the \ac{LVK} collaboration has already reported 90 events involving the merger of stellar mass compact binaries
\cite{abbott2021gwtc,abbott3ligo,scientific2021gwtc,abbott2019tests},
which include \ac{BBH}, \ac{BNS}, and \ac{NSBH}
\cite{abbott2019tests,broekgaarden2021formation}.
Besides \acp{BH} and \acp{NS}, some models of exotic objects \cite{johnson2020constraining} such as quarkstars\cite{haensel1986strange}, \acp{BS} \cite{kolb1993axion}, gravastars\cite{mazur2023gravitational}, and \acp{BH} in modified theories of gravity
are also proposed as alternatives\cite{cardoso2019testing}.
The \acp{GW} generated by the bianries constituted of these \acp{ECO} will differ from those produced by \ac{BBH}.
Therefore, we can use \ac{GW} to test the nature of the compact objects.

According to the black hole no-hair theorem \cite{hawking2005information}, the classical black holes in general relativity are fully characterized by their masses, spins, and charges.
However, due to various neutralization mechanisms \cite{blandford1977electromagnetic},
it’s widely believed that astrophysical \acp{BH} will have negligible electric charge.
So these \acp{BH} can be characterized by the Kerr metric, which includes only the mass $M$ and the spin $a$ as the parameters.
By measuring multiple parameters of a \ac{BH}, and testing if they could provide a consistent prediction of $M$ and $a$ according to \ac{GR}, we can test the no-hair theorem and probe the nature of the compact objects.

Various parametrization methods have been proposed for such tests,
including the tidal deformability \cite{hinderer2010tidal,chatziioannou2020neutron,raithel2018tidal,malik2018gw170817},
the horizon absorption effect\cite{bernuzzi2012horizon}, the quasi-normal mode spectrum of ringdown \cite{kokkotas1999quasi,kelly2013decoding,ota2020overtones,baibhav2018black,shi2019science,berti2006gravitational,berti2007matched,dreyer2004black,detweiler1980black}
and the multipole moments \cite{poisson1998gravitational,kastha2018testing}.
With these parameterizations, the \ac{BH} will perform differently compared to the mimickers.

For a localized object, its gravitational field can be expanded in terms of the multipole moments \cite{hansen1974multipole,geroch1970multipole,thorne1980multipole,compere2018gravitational}.
For stationary asymptotically flat solutions of the Einstein equation,
such as the Kerr black hole, the multipole moments can be expressed by the mass $M$ and spin $a$ as
\begin{equation}\label{multipole}
\mathcal{M}_l+{\rm i}\mathcal{S}_l=M({\rm i}a)^l
\end{equation}
There are two sets of multipole moments, the mass moments $\mathcal{M}_l$ for even $l$s, and the current moments $\mathcal{S}_l$ for odd $l$s.
The mass multipole moments for odd orders and the current multipole moments for even orders will vanish due to the equatorial symmetry of the Kerr solution.
The leading-order mass moment $\mathcal{M}_0=M$ and current moment $\mathcal{S}_1=Ma$ are the mass and spin angular momentum of the Kerr \ac{BH}, respectively.
If we can measure the multipole moments with $l\geq2$ besides the mass and spin, then we can test if these expression are broken, and thus test the no-hair theorem.

In most cases, only the $l=2$ term, known as the \ac{SIQM}, is considered in the relevant test.
For a general compact object, the \ac{SIQM} can be represented as $Q=-\kappa\chi^2{M^3}$ where $\chi=a/M$ is the dimensionless spin parameter, and $\kappa$ is a coefficient that depends on the internal structure of the object related to its equation of state \cite{krishnendu2017testing}.
For \acp{BH}, we will have $\kappa=1$ according to \eqref{multipole}.
For \acp{NS}, it's belived that $\kappa$ can vary between 2 and 14 \cite{laarakkers1999quadrupole,pappas2012revising} due to the multipole deformation that occurs during the rotation process\cite{stein2014three}, up to quadratic in spin.
For \acp{BS}, the range of $\kappa$ is about 10 to 150 \cite{herdeiro2014kerr,baumann2019probing}.
For some other BH mimickers such as gravastars, the value can also be negative\cite{uchikata2016tidal,dubeibe2007chaotic}.
By measuring $\kappa$, we can distinguish between the \acp{BH} and its mimickers\cite{ryan1997accuracy,ryan1997spinning}.

Using the low-mass events in GWTC-2\cite{narikawa2021gravitational}, the data support the model of \ac{BBH} rather than \ac{ECO}, and $\kappa$ is constrained to the order of $\mathcal{O}(10^2)$.
Recent work has also analyzed the impact of spin precession and higher modes on the measurement of \ac{SIQM} \cite{Divyajyoti2023effect}, and the constraint on octupole moments \cite{Saini:2023gaw} with ground-based detectors. Some selected GWTC events are also used in the data analysis.
The combined Bayesian factor among the GWTC events is calculated, $\rm logBF_{\delta\kappa_s\neq0}^{\rm Kerr}=0.9$ \cite{abbott2021tests} in GWTC-3 and 1.1 in GWTC-2\cite{abbott2021tests2}.
The capability is also analyzed for  LISA and DECIGO \cite{krishnendu2020testing} with the detection of \ac{MBHB},
and $\kappa$ is expected to be constrained to the order of $\mathcal{O}(0.1)$.
Based on some astrophysical models for the population of \ac{MBHB}, it's also argued that 3\% of the events can reach these levels.
Moreover, with the detection of \acp{EMRI}, TianQin\cite{fan2020science,zi2021science,liu2020augmented} and LISA\cite{barack2004lisa,barack2007using,babak2017science,Rahman:2021eay} can constrain the \ac{SIQM} to $10^{-5}$.

TianQin is a space-borne \ac{GW} dectector \cite{luo2016tianqin,hu2017science} to be launched in 2035.
It comprises three drag-free satellites orbiting the Earth at radius of $10^5$ km and aims to dectect \acp{GW} on mHz band.
The major objectives \cite{mei2021tianqin} include the merger of \acp{MBHB} \cite{wang2019science,feng2019preliminary},
the inspiral of stellar-mass \acp{BBH} \cite{liu2020science,liu2022capability},
the of galactic compact binaries \cite{huang2020science},
the \acp{EMRI}\cite{fan2020science,Fan2022emrb},
and the stochastic \ac{GW} background \cite{liang2023sensitivity,liang2022science}.
With the observation of these signals, we can also study the evolution of the universe \cite{Zhu:2021aat,Zhu:2021bpp,Huang:2023prq},
and the nature of \acp{BH} and gravity
\cite{shi2019science,bao2019constraining,zi2021science,Sun:2022pvh,Xie:2022wkx,Shi:2022qno,Lin:2023ccz}.

In this work, we conduct a more comprehensive study on the effectiveness of TianQin in testing NHT by probing the \ac{SIQM} with the inspral signal of \acp{MBHB}.
According to the result of \cite{wang2019science}, TianQin is expected to detect abotu 60 events every year for the most optimistic model.
We consider the higher $(l,m)$ mode corrections due to the deviation of the \ac{SIQM}, and utilize \ac{TDI} response to generate the signal.
With the \ac{FIM} analysis, we find that the late inspiral will dominate the accuracy of the constraint.
This indicates that we do not need to consider a full inspiral signal in this analysis.
The results also indicate that events with asymmetric mass will have better capability, and higher modes will be important for events with large mass ratios.
So, we also consider these higher modes in our waveform and the corresponding modifications.
Then we use \texttt{bilby} to conduct the Bayesian analysis, and the accuracy of the parameter estimation is consistent with the \ac{FIM} result.
For the injection signal with a non-zero $\delta\kappa$, if we do not consider this deviation in the matched waveform, the result will exhibit a significant bias in the estimation of other source parameters.
By calculating the Bayes factor, we find that the signal from \acp{BH} and \acp{ECO} can be distinguished without a doubt.

The paper is organized as follows.
In Section \ref{sec:method}, we will provide a brief review of the basic methods for waveforms, responses, and statistics in each subsections,respectively. Then, we present our results for TianQin with \ac{FIM} and Bayesian analysis in Section \ref{sec:result}.
Finally, we provide a brief summary of conclusion in Section \ref{sec:con}.
Throughout this work, the geometrized unit system $(G=c=1)$ is used.

\section{Method}\label{sec:method}

\subsection{Waveform}

In this work, we utilize the \texttt{IMRPhenomXHM} \cite{garcia2020multimode} waveform,
which is a frequency domain model for the inspiral-merger-ringdown of quasi-circular non-precessing \ac{BBH} with higher modes.
In general, the waveform can be expressed as
\begin{equation}
h_{BBH}(f)=\sum_{lm}A_{lm}(f)e^{i\Psi_{lm}^{BBH}(f)}.
\end{equation}
$A_{lm}(f)$ and $\Psi_{lm}^{BBH}(f)$ are the amplitude and phase for the $lm$ mode, respectively.
The index `BBH' ndicates that the corresponding formula is derived for \acp{BBH}, and it will be different for binaries constitude of \acp{ECO}.
The \ac{SIQM} of the progenitors will influence the phase evolution of inspiral, while the remnant \ac{SIQM} will affect the quasi-normal mode spectrum of ringdown.
In this work, we will focus on the inspiral phase, and therefore a cutoff at the \ac{ISCO} will be adopted in the subsequent calculations.
Since the spin precession is not considered in the waveform we used,
we will assume the aligned or anti-aligned spin for the binaries.

For binaries constitude of \acp{ECO}, the waveform for inspiral can be modified.
\begin{equation}
h_{ECO}(f)=\sum_{lm}A_{lm}(f)e^{i\left(\Psi_{lm}^{BBH}(f)+\Psi_{lm}^{SIQM}(f)\right)},
\end{equation}
This implies that we disregard the modification of the amplitude\cite{arun2009higher}because the phase will have greater impact on the accuracy of \ac{PE} for intrinsic parameters.
If we neglect the tidal effect and only consider the leading-order correction of SIQM,
the phase correction for the leading $22$ mode can be expressed as
\cite{narikawa2021gravitational,krishnendu2017testing,arun2009higher,mishra2016ready}
\begin{equation}
\Psi^{SIQM}_{22}(f)=\frac{75}{64}\frac{\delta\kappa_1M^2_1\chi^2_1+\delta\kappa_2M^2_2\chi^2_2}{M_1M_2}(\pi M_{tot}f)^{-\frac{1}{3}}. \label{con:eq0}
\end{equation}
$M_{tot}=M_1+M_2$ is the total mass of the binary system,
and $\delta\kappa_i=\kappa_i-1$ characterizes the deviation of the \ac{SIQM} relative to the \ac{BH}.
It should be noted that the mass we used is the redshifted mass all through this paper.
The power index of $-1/3$ means that this leading-order correction emerges at the 2PN order.
We set $\delta\kappa_1=\delta\kappa_2=\delta\kappa$, and thus ignore the antisymmetric contribution.
Obviously, the \ac{BBH} cases correspond to $\delta\kappa=0$, and we have neglected the \ac{BH}-\ac{ECO} system or binary \acp{ECO} with different $\kappa$.

For the correction of the phase of the higher modes, we utilize the relation provided in the parameterized post-Einsteinian framework \cite{mezzasoma2022theory}.
Since the leading-order correction emerges at 2PN order, the higher modes correction can be written as:
\begin{equation}\label{con:eq1}
\psi_{lm}^{SIQM}=\left(\frac{2}{m}\right)^{-4/3}\psi_{22}^{SIQM}.
\end{equation}
In our analysis, the modes we considered include the dominant mode (2,2) and subdominant modes (2,1) and (3,3).

\subsection{The Response and Noise of TianQin}\label{subsec:detector}

For space-borne \ac{GW} detectors, \ac{TDI} \cite{Tinto:2003vj,marsat2021exploring} must be used to suppress the laser phase noise.
In this work, we utilized the 1.0 type channel for $A$, $E$, $T$ for data analysis.
The signal is obtained by multiplying the waveform with the transfer function.
\begin{equation}
h_{\Tilde{A},\Tilde{E},\Tilde{T}}=\sum_{lm}\mathcal{T}^{lm}_{\Tilde{A},\Tilde{E},\Tilde{T}}\Tilde{h}_{lm}(f)
\end{equation}
For details on the formalism of the transfer function for TianQin, please refer to \cite{lyu2023parameter}.
\textcolor{red}{The orbit motion of TianQin is considered as the ideal case described in \cite{luo2016tianqin}.
Thus the modulation caused by the rotation of the constellation around the earth, and the Doppler effect caused by the motion of the constellation around the sun will be included in the calculation of the response.}
In this work, we only consider the $A$ channel in our calculations.

The \ac{PSD} for the noise of TianQin corresponding to the \ac{TDI} channel can be written as
\begin{equation}
\begin{split}
S_{A,E}&=8\sin^2\frac{f}{f_*}\left[(2+\cos\frac{f}{f_*})(2\pi f)^2S_x\right.\\
&\left.+4\left(1+\cos\frac{f}{f_*}+\cos^2\frac{f}{f_*}\right)\left(1+\frac{0.1\rm{mHz}}{f}\right)\frac{S_a}{(2\pi f)^2}\right]
\end{split}
\end{equation}
where $f_*=\frac{1}{2\pi{L}}$ is the characteristic frequency of TianQin,
and $L=\sqrt{3}\times10^{8}$m is the arm length.
The acceleration noise $S_a$ and the position noise $S_x$ is\cite{luo2016tianqin}

\begin{equation}
\begin{split}
&S_a= 10^{-30}{\rm m}^2\cdot{{\rm s}^{-4}}\cdot{{\rm Hz}^{-1}}\\
&S_x= 10^{-24}{\rm m}^2\cdot{{\rm Hz}^{-1}}
\end{split}
\end{equation}

\subsection{Parameter Estimation}

In this study, \ac{FIM}  is used to estimate TianQin's capability to measure the parameters.
By the definition of an inner product
\begin{equation}
(g|h)\equiv2\int_{f_{low}}^{f_{high}}\frac{{g(f)^*}h(f)+{h(f)^*}g(f)}{S_n(f)}df,
\end{equation}
The \ac{FIM} is defined as \cite{cutler1994gravitational}
\begin{equation}
\Gamma_{ij}\equiv\left(\frac{\partial{h}}{\partial\theta_i}|\frac{\partial{h}}{\partial\theta_j}\right)
\end{equation}
$h$ is the response signal of the injected waveform, and $\theta_i$ is the $i$-th parameter of the source.
In our analysis, the parameters are chosen as follows
\begin{equation}\label{eq:para}
\theta = \left\{{\mathcal{M}, \eta, D_L,t_c,\phi_c, \psi, \iota, \chi_1, \chi_2, \delta\kappa_s  }\right\}
\end{equation}
$\eta=\frac{M_1M_2}{(M_1+M_2)^2}$ is the symmetric mass ratio, $\mathcal{M}=M_{tot}{\eta}^{3/5}$ is the chirp mass, and $D_L$ is the luminosity distance.
$t_c$ and $\phi_c$ represent the time and phase at coalescence.
$\psi$ and $\iota$ represent the polarization and inclination angles of the source.
$\chi_1$ and $\chi_2$ represent the dimensionless spin parameters for each \acp{BH}.
For a signal with a large \ac{SNR}, the uncertainty in
parameter estimation is given by
$$\Delta\theta_i=\sqrt{\Gamma^{-1}_{ii}}$$

In the calculation of the inner product, the lower and higher frequency bounds are chosen as
\begin{equation}
\begin{split}
&f_{high}=min(f_{max},f_{isco})\\
&f_{low}=max(f_{min},f_{init})
\label{con:eq3}
\end{split}
\end{equation}
Due to the sensitivity band of TianQin, we set $f_{max}$=1Hz, $f_{min}$=$10^{-4}$Hz,
and signals outside of this band will be ignored.
The initial frequency of the source,$f_{init}$ depends on the time $T$ we begin to observe before the merger of the \ac{BBH}
\begin{equation}
f_{init}=(\frac{5}{256})^{3/8}\frac{\mathcal{M}^{-5/8}}{\pi}{T}^{-3/8}
\end{equation}
$f_{isco}$ is the frequency for the \ac{ISCO}, which marks the end of the inspiral.
In our calculation, we use the Kerr frequency  instead of Schwarzschild because spin plays a crucial role in our calculation, and it significantly affects the radius of \ac{ISCO}.
The detailed formalism can be found in Appendix ~\ref{app:ISCO}.

For a more realistic analysis, we also use Bayesian inference\cite{thrane2019introduction} to perform \ac{PE} for simulated data.
In the Bayesian framework, the posterior distribution $p(\theta|D)$ for a specific set of parameters $\theta$ with the given data $D$ is
\begin{equation}
p(\theta|D) = \frac{p(D|\theta)p(\theta)}{p(D)}
\end{equation}
In the equation above, $p(\theta)$ represents the prior, $p(D|\theta)=\mathcal{L}$ represents the likelihood,
and $p(D)=\mathcal{Z}$ is the evidence necessary to ensure the normalization condition of the posterior
\begin{equation}
\int{p(\theta|d)d\theta}=1
\end{equation}
For GW detection, we typically assume that the noise is stationary and Gaussian. In this case, the likelihood can be expressed as:
\begin{equation}
\ln\mathcal{L}\propto{-\frac{1}{2}(D-h(\theta)|D-h(\theta))}
\end{equation}
$h(\theta)$ is the waveform template for a given set of parameters $\theta$.
The proportionality coefficient is not relevant to $\theta$.
The evidence is then defined as
\begin{equation}
\mathcal{Z}=\int{\mathcal{L}(D|\theta)p(\theta)}d\theta
\label{con:eq2}
\end{equation}

Beyond the calculation of the posterior, we can also investigate model selection,
which involves determining which model is favored by the observed data.
This can be achieved by calculating the Bayes factor between two models $\mathcal{M}_1$ and $\mathcal{M}_2$:
\begin{equation}
\rm{BF}_2^1=\frac{\mathcal{Z}_1}{\mathcal{Z}_2},
\end{equation}
and $\mathcal{Z}_i$ is the evidence of the model $\mathcal{M}_i$.
The $\lg$ Bayes factor is most commonly used, and it is defined as:
\begin{equation}\label{equ:bf}
\lg \rm{BF}_2^1=\lg \mathcal{Z}_1-\lg\mathcal{Z}_2,
\end{equation}

In the calculation of the posterior distribution, we use \texttt{bilby} \cite{ashton2019bilby} to implement parameter estimation, which is primarily designed for inferring compact binary coalescence events from interferometirc data.
For sampling across the parameter space, we utilized \texttt{dynesty} \cite{speagle2020dynesty} based on the nested sampling algorithm \cite{skilling2006nested} .
However, since \texttt{bilby} is designed for the ground-based detectors,
we modified the components related to the response and noise of the detector,
as we introduce in Section \ref{subsec:detector}.

\section{Result}\label{sec:result}

The default parameters of the \acp{MBHB} we used for both Fisher and Bayesian analysis are shown in Table \ref{tab:table1}.
We also list the prior we use in the Bayesian analyses for each parameter in this table.
Beside the parameters listed in \eqref{eq:para} ,
$\beta$ and $\lambda$ represent the latitude and longitude for the sources in ecliptic coordinates.
Since we only consider the 1 day data in our analysis, the detector's response will not change significantly.
Thus, the position of the source that will influence the response is poorly constrained.
So, we estimate all the parameters in Table \ref{tab:table1} except $\beta$ and $\lambda$.

\begin{table}[htbp]
\caption{
The default values and the prior of the parameters we choose for the \acp{MBHB}.
}
\begin{ruledtabular}
\begin{tabular}{ccc}
\textrm{Parameter}&\textrm{Value}&\textrm{Prior}\\
\colrule
$\mathcal{M}$($M_{\odot}$) & $1.24\times10^6$ &\rm logarithm uniform $[10^2,10^8]$\\
$\eta$ & $\frac{2}{9}$&\rm uniform$[\frac{1}{12},\frac{1}{4}]$\\
$D_L$(Mpc) & 1000&\rm quadratic uniform$[10^2,10^4]$\\
$t_c$(s) & 3600&\rm uniform$[-10000,10000]$\\
$\phi_c$(rad) & $\frac{\pi}{4}$&\rm uniform$[-\pi,\pi]$\\
$\psi$(rad) & $\frac{\pi}{4}$&\rm uniform$[-\pi,\pi]$\\
$\iota$(rad) & $\frac{\pi}{4}$&\rm cosine uniform$[0,\pi]$\\
$\chi_1$ & 0.2&\rm uniform$[-1,1]$\\
$\chi_2$ & 0.1&\rm uniform$[-1,1]$\\
$\beta$(rad) & $\frac{\pi}{4}$&N/A\\
$\lambda$(rad) & $\frac{\pi}{4}$&N/A\\
$\delta\kappa$ & $0$ & \rm uniform$[-20,20]$
\end{tabular}
\end{ruledtabular}
\label{tab:table1}
\end{table}

\subsection{Fisher Analysis on the Capability of TianQin}

In the sensitive frequency band of TianQin, the inspiral of the \acp{MBHB} may last for years before merging.
However, it has been found that the late inspiral part will capture most of the \ac{SNR} \cite{feng2019preliminary}.
According to the results shown in Table \ref{tab:table2},
we can observe that the signal from the last day contributes $99.9\%$ of the \ac{SNR} for the entire year.
The \ac{PE} accuracy for 1 month is almost equivalent to the accuracy for 1 year,
whereas the accuracy for 1 day is approximately 1.8 times lower than the accuracy for 1 year.
So, in the Bayesian analysis which will be discussed in the following subsection,
we will only consider the analysis of the data from the final day before the merger
to reduce the cost of  computation.
More over, according to the Fisher analysis with or without $\beta$ and $\lambda$,
the accuracies for the constraint on $\delta \kappa$ are almost the same,
and the correlation between $\delta \kappa$ and the sky position is very weak.
However, the \ac{PE} for the position of the source is determined by the modulation of the response function which caused by the movement of the  detector.
If we only consider the data from the last day, the detector will not move significantly,
and thus the estimate of the position will be very worse.

This can be easily solved by considering a longer data.
Since we only focus on the estimation of $\delta\kappa$,
and it will not have a correlation with the latitude and longitude of the source,
we will ignore these two parameters in the following analysis.

\begin{table}[htbp]
\caption{
The \ac{SNR} and \ac{PE} accuracy for the default source with varying duration.}
\begin{ruledtabular}
\begin{tabular}{lcdr}
\textrm{Duration Time}&
\textrm{SNR}&
\delta\kappa\\
\colrule
1 year & 6641.35&0.174\\
1 month & 6641.35&0.174\\
1 week & 6641.16&0.213\\
1 day &6639.58&0.328\\
\end{tabular}
\end{ruledtabular}
\label{tab:table2}
\end{table}

\begin{figure}[htbp]
\includegraphics[width=1\columnwidth]{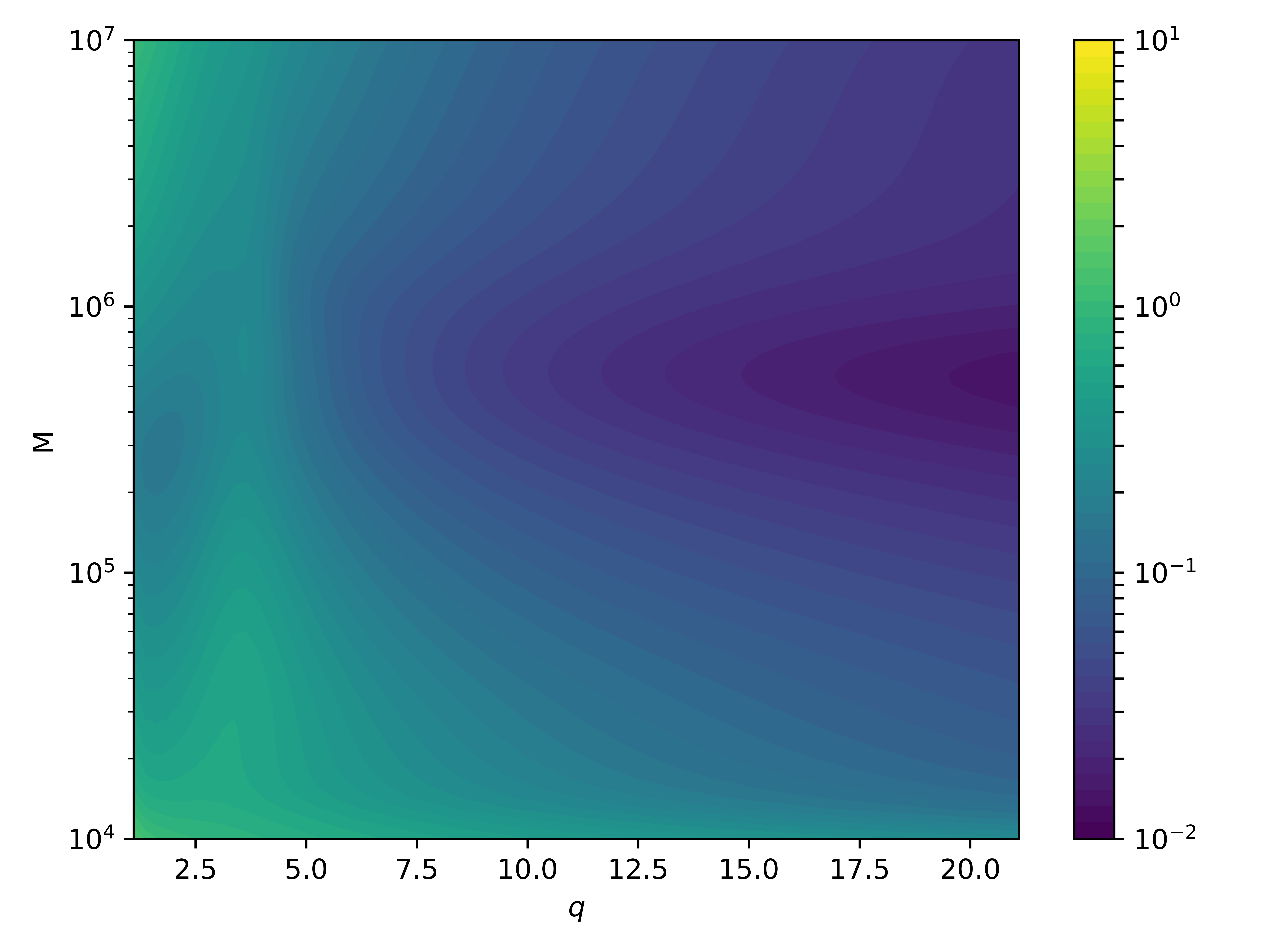}
\caption{
The contour plot shows the \ac{PE} accuracy of $\delta\kappa$ for the sources with different total mass and mass ratio values over a signal duration of 1 day.
The \ac{SNR} for each point is normalized to 5000 by changing $D_L$.}
\label{fig:epsart2}
\end{figure}

According to the results above, we can see that TianQin can constrain the \ac{SIQM} to the level of $\mathcal{O}(0.1)$.
Then, we calculate the capacity for the sources with different total mass and mass ratio.
The result is shown in Fig \ref{fig:epsart2} as a contour plot.
For a fair comparison, the \ac{SNR} is normalized to 5000 by adjusting the luminosity distance $D_L$.
The mass ratio $q$ varies between $1$ and $21$, and the total mass varies between $10^4M_\odot$ and $10^7M_\odot$.

According to the contour plot, we can see that TianQin has better sensitivity for sources around $10^{5.5}M_\odot$.
This region corresponds to the most sensitive band of TianQin.
Comparing this with the result of LISA obtained in \cite{krishnendu2020testing},
we can see that LISA has better sensitivity for sources around $2\times10^{6}M_\odot$,
since it's more sensitive for the lower frequencies than TianQin.
Similar phenomenons have also been found for other studies of the detection of \acp{MBHB},
and more detailed comparison between LISA and TianQin can be found in \cite{Torres-Orjuela:2023hfd}. 
It also shows that events with a higher mass ratio will provide better constraints.
This is consistent with the result of \ac{EMRI} \cite{zi2021science},where the mass ratio become $10^6$ and the constraint of the \ac{SIQM} reaches the level of $10^{-4}$.
For events with asymmetric masses, the higher modes will become important.
According to previous studies, introducing modifications on the higher modes can also enhance the capability  \cite{Wang:2023wgv}.
This will be discussed with Bayesian analysis in the next subsection.

\subsection{Bayesian analysis}

\begin{figure}[htbp]
\centering
\includegraphics[width=1\columnwidth]{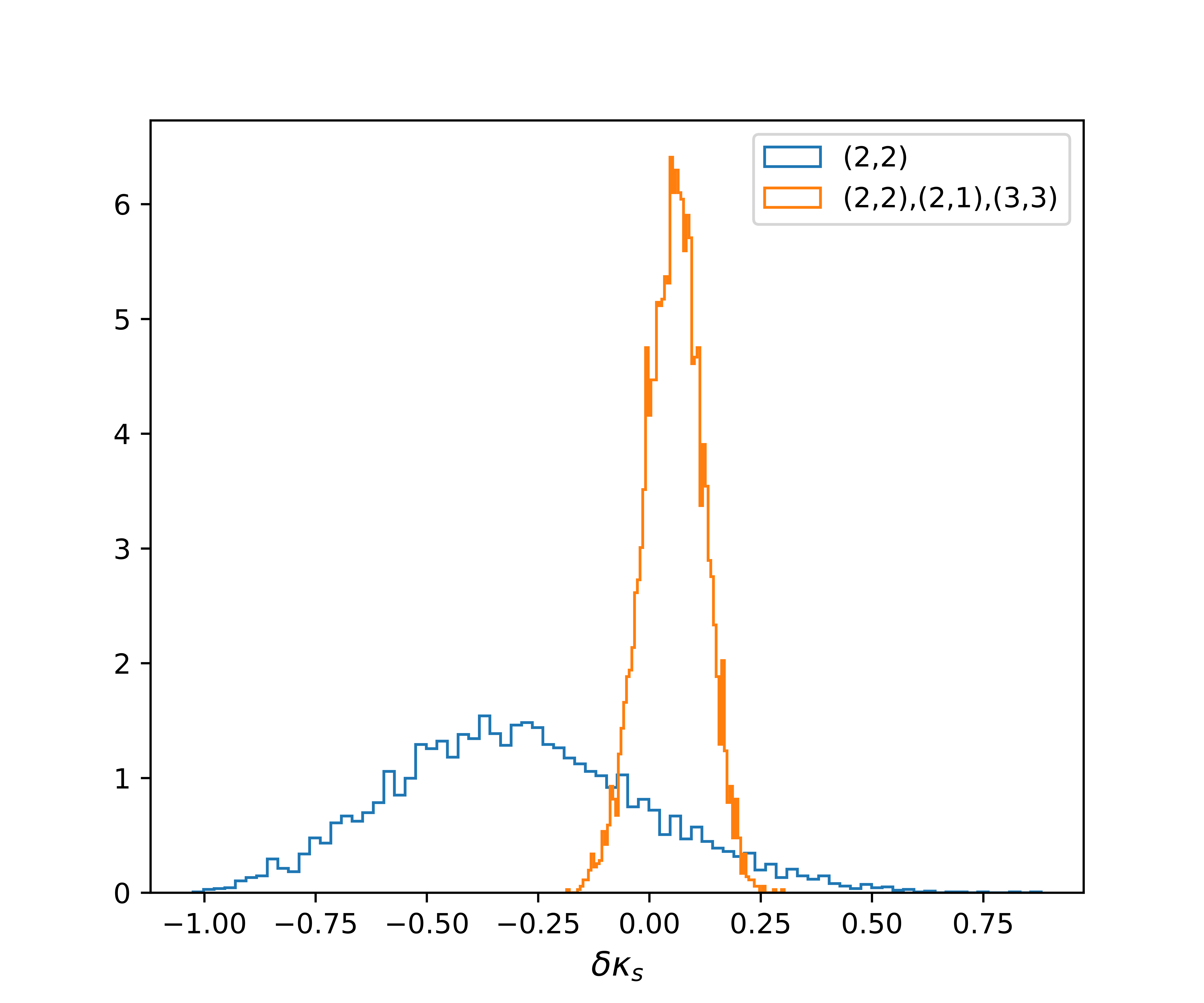}
\includegraphics[width=1\columnwidth]{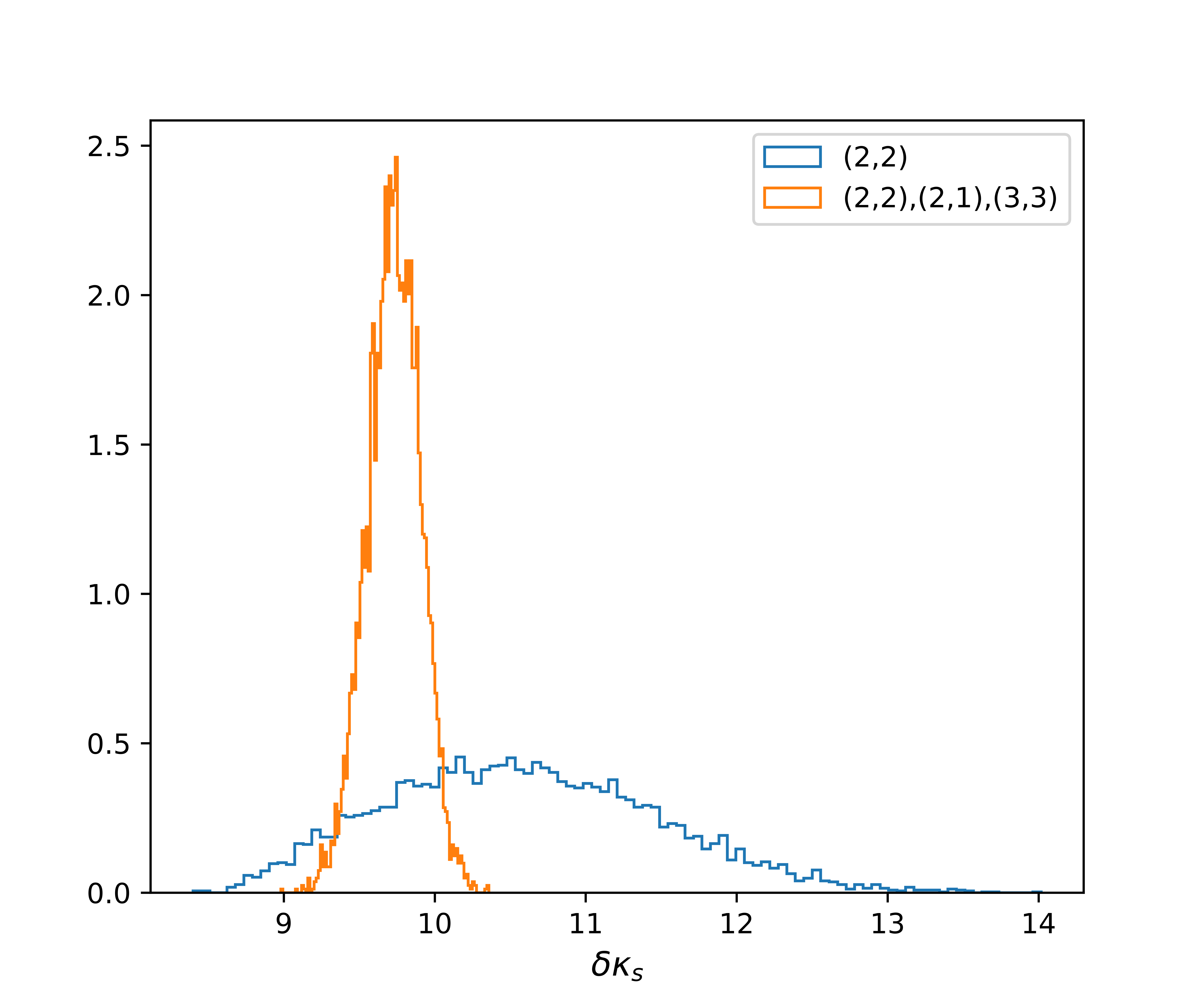}
\caption{
The \ac{PDF} of $\delta\kappa_s$ for the waveform with the $(2,2)$ mode only (blue) and with higher modes (orange).
The top panel shows the result for the injected signal with higher modes and $\delta\kappa_s=0$,
The bottom panel shows the result for the injected signal with higher modes and $\delta\kappa_s=10$.}
\label{fig:epsart5}
\end{figure}

\begin{figure*}[htbp]
\centering
\includegraphics[width=1\textwidth]{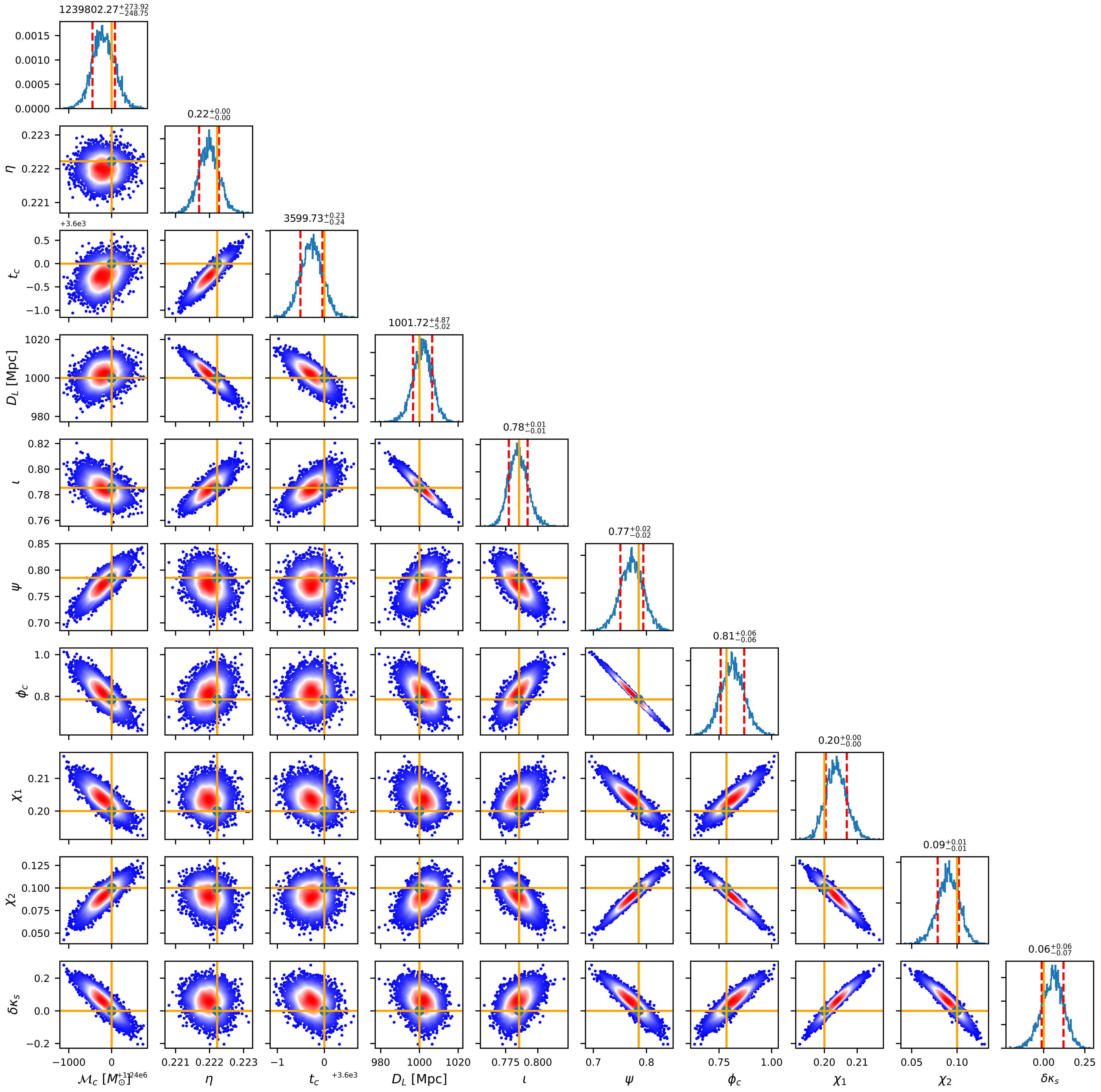}
\caption{
The \ac{PE} result with higher modes included waveform,
for the injected signal with higher modes and $\delta\kappa_s=0$.
The \ac{SIQM} is included in the estimated parameters.
}
\label{fig:epsart6}
\end{figure*}

For the Bayesian analysis, we consider two types of injected data:
the first one is the \ac{BBH} signal with $\delta\kappa=0$,
and the second one is the binary \ac{ECO} signal with $\delta\kappa=10$.
It should be noted that the mock data in our analysis is an idealized case.
In the real data for space-borne \ac{GW} detectors, there may exist multiple different kinds of signals at the same time.
Thus a global fit method \cite{Littenberg:2023xpl,Katz:2024oqg} must be used in the pipeline of data analysis.
More over, we also assumed that the signal of \ac{MBHB} has already been detected with the search pipeline such as \cite{Chen:2023qga},
and all the analysis we did in this work is just parameter estimation \cite{Gao:2024uqc} and model selection.
Both injections have three primary modes: $(2,2)$, $(2,1)$, and $(3,3)$, with all the parameters selected according to Table \ref{tab:table1}.
For the waveform used in the matched filtering, we did not assume the model of \ac{BBH},
which means that $\delta\kappa$ is also a parameter that needs to be estimated and can not be set as an constant.
The marginalized \acp{PDF} are shown in Fig \ref{fig:epsart5}.
The results show that by incorporating the higher modes in the waveform,
the capability will be improved for about 3 times.
The \ac{PE} result for all parameters with \ac{BBH} injection is illustrated in Fig \ref{fig:epsart6} as an example.
We can see that all the parameters are estimated properly,
which indicated that all true values are contained in the $1-\sigma$ region.

\begin{figure*}[htbp]
\includegraphics[width=1\textwidth]{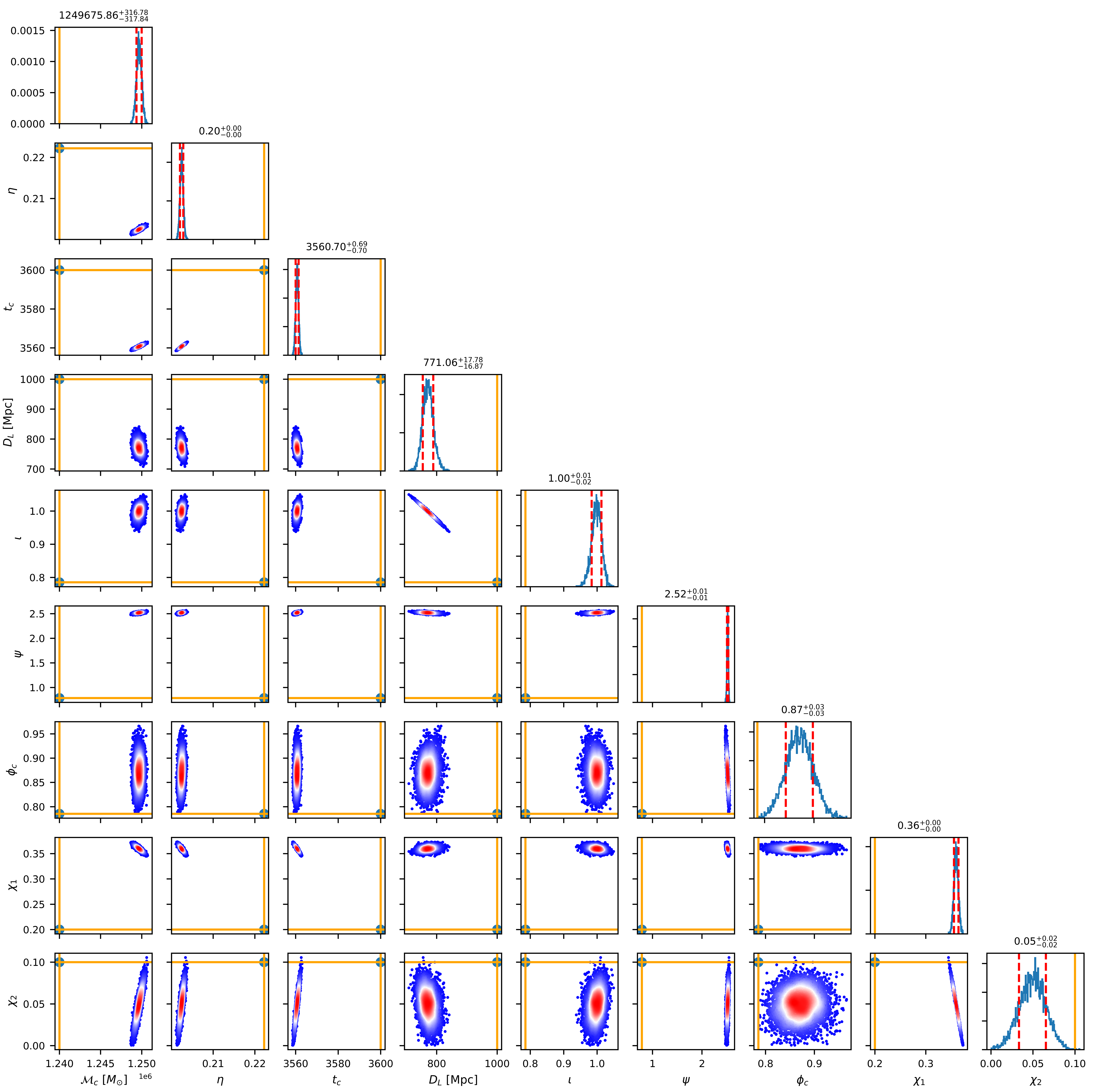}
\caption{\label{fig:epsart7}
This figure demonstrates that the corner plots of the waveform template are consistent with $\delta\kappa_s=0$ but the injected signal has $\delta\kappa_s=10$ with 3 modes considered. We can easily notice that the estimated central values deviate from the injected values. This implies that our hypothesis in this figure is incorrect.
 }
\end{figure*}

We also analyzed the case of a non-\ac{BBH} injection with the estimation of \ac{BBH} waveform.
The injected value is $\delta\kappa=10$, and it is fixed to be 0 in the Bayesian analysis.
This corresponds to a scenario where the data is generated by a binary \ac{BS} system, but we incorrectly use a \ac{BBH} waveform to analyze it.
The \ac{PE} result is shown in Fig \ref{fig:epsart7}.
We can see that all the parameters are estimated with a significant bias.
For example, the injected value of the chirp mass is $1.24\times10^6M_\odot$,
but the estimated result is $1.25\times10^6~^{+160}_{-163} M_\odot $,
the true value is approximately $60-\sigma$ away from the point with the highest likelihood.
This means that if we use the \ac{BBH} waveform to fit a binary \ac{ECO} signal, the parameters we estimate will deviate significantly from the real values.

We also calculate the Bayes factors for the \ac{BH} hypothesis compared to the \ac{ECO} hypothesis.
Here, we analyzed two types of injections: one with $\delta\kappa_s=0$ representing the \ac{BH}, the other with $\delta\kappa_s=10$ representing the \ac{BS}.
The Bayes factor is calculated according to \eqref{equ:bf} for Model 1 corresponding to \ac{BH} and Model 2 corresponding to \ac{ECO}.
Both injections include the higher modes. We consider the case of estimation with only the $(2,2)$ mode and the case where higher modes are included.

\begin{table}[htbp]
\caption{This table displays the $\lg \rm BF^{\rm BH}_{\rm ECO}$ with an injected signal where $\delta\kappa_s$=0 or 10, with or without consideration of higher modes.
}
\begin{ruledtabular}
\begin{tabular}{lccr}
\textrm{injected $\delta\kappa_s$}&
\textrm{(2,2) mode only}&
\textrm{include higher modes}\\
\colrule
0& 0.19& 1.49\\
10& -188.05&-4740.35\\
\end{tabular}
\end{ruledtabular}
\label{tab:table3}
\end{table}

For the injection of \ac{BH}, the results will support the \ac{BH} hypothesis,
but the Bayes factor is not very large.
$\rm log{\rm BF^{\rm BBH}_{\rm ECO}}$ will be 0.19 if we only consider the (2,2) mode,
and the support for the true model is weak.
But the result will increase to 1.49 if we consider the higher modes, and the support for the true model is strong.
For the injection of \ac{ECO}, the results will strongly support the \ac{ECO} hypothesis,
and the Bayes factor will become very large.
$\rm log{\rm BF^{\rm BBH}_{\rm ECO}}$ will be -188.05 if we only consider the (2,2) mode,
and it will become -4740.35 if we consider the higher modes.
Both results will support the true model with a very strong evidence.
This means that we can distinguish between the \ac{BH} and \ac{ECO} models by calculating the Bayes factor, which can help us avoid potential systematic errors.

\section{Conclusion}\label{sec:con}

The no-hair theorem states that the multiple moments of a \ac{BH} are entirely determined by its mass and spin, and it will be violated for \acp{ECO}.
This work focuses on testing the no-hair theorem by probing the \ac{SIQM} of the \acp{BH} using the inspiral signal of a \ac{BBH} system.
We consider the space-based \ac{GW} detector TianQin as an example, and then the source chosen is the \acp{MBHB}.

With the analysis using the Fisher matrix, we find that TianQin has the best capability for sources with a total mass around $10^{5.5}M_\odot$, corresponding to the sensitive band of TianQin.   For LISA\cite{krishnendu2020testing}, the best capability total mass is around $2\times10^{6}M_\odot$ and both can constrain their appropriate \acp{MBHB} sources'  \acp{SIQM} to $\mathcal{O}(10^{-1})$ order.
Our results also show that \acp{BBH} with a larger mass ratio will have better constraints, indicating the need to consider higher modes.

Then we conducted the analysis using Bayesian inference.
The result agrees with the estimation using the Fisher matrix for both the \ac{BH} and \ac{ECO}models.
The accuracy will improve by about 3 times if we include the higher modes.
When using the \ac{BH} model to infer the parameters of a binary \ac{ECO} system, we also observe that the estimation of the parameters will have significant systematic errors.
However, this can be avoided by  calculating the Bayes factor,which will provide strong evidence to distinguish between different models.

As a preliminary exploration, our work still has some limitations.
For example, in the model with a non-zero $\delta\kappa$, we have assumed that both \acp{ECO} in the binary system have the same value of $\delta\kappa$, and thus $\delta\kappa_a$ is fixed at zero.
Obviously, this could not be the case in the real world, but the degeneracy between the parameters restricts us from estimating $\delta\kappa_s$ and $\delta\kappa_a$ simultaneously.
Moreover, we use the data for only 1 day to perform the \ac{PE} to reduce the computation.
Although we have proven that this does not compromise the generality, and the results will not vary significantly for longer datasets.
But this is not the case for real data analysis, and the position of the source cannot be estimated in this scenario.
We leave the inclusion and treatment of these more realistic issues for future exploration.

\begin{acknowledgments}\label{sec:acknowledgments}

The authors thank Yi-Ming Hu, Jiangjin Zheng, Xiangyu Lv and MengKe Ning for their helpful discussion.
We acknowledge the usage of the calculation utilities of bilby \cite{ashton2019bilby},  NUMPY \cite{vanderWalt:2011bqk}, and SCIPY \cite{Virtanen:2019joe}, and plotting utilities of MATPLOTLIB \cite{Hunter:2007ouj} and Corner \cite{Foreman-Mackey2016}.
This work is supported by the Guangdong Basic and Applied Basic Research Foundation (Grant No. 2023A1515030116), the Guangdong Major Project of Basic and Applied Basic Research (Grant No. 2019B030302001), and the National Science Foundation of China (Grant No. 12261131504).
This work uses bilby

\end{acknowledgments}

\appendix

\section{The frequency for ISCO}\label{app:ISCO}

In this appendix we present the formula corresponding the the \ac{ISCO} frequency for \ac{BBH} with spin \cite{favata2022constraining}.
The ISCO frequency can be written as
\begin{equation}
f_{\rm ISCO}=\frac{\hat{\Omega}(\chi_f)}{\pi{M_f}},
\end{equation}
where $\chi_f$ represents the final spin, and $M_f$ represents the final mass of the remnant \ac{BH} after the merger of \ac{BBH}.
The detailed calculation can be found in references \cite{husa2016frequency,hofmann2016final}.
$\hat{\Omega}(\chi_f)$ represents the dimensionless ISCO Kerr angular frequency.
\begin{equation}
\hat{\Omega}(\chi_f)=\frac{1}{{\hat{r}_{\rm ISCO}}^{3/2}(\chi_f)+\chi_f}
\end{equation}
The dimensionless radius of the \ac{ISCO} for a Kerr \ac{BH}  with a dimensionless spin parameter $\chi$ is
\begin{equation}
\begin{split}
\hat{r}_{\rm ISCO}(\chi)=3+Z_2&-\frac{\chi}{|\chi|}\sqrt{(3-Z_1)(3+Z_1+2Z_2)}\\
&Z_2=\sqrt{3\chi^2+Z_1^2}\\
Z_1=1+(1-\chi^2)&^{1/3}((1+\chi)^{1/3}+(1-\chi)^{1/3})
\end{split}
\end{equation}

\bibliography{SIQM}
\end{document}